\documentclass[a4paper,12pt]{article}

\makeatletter

\@addtoreset{equation}{section}
\makeatother
\usepackage{amsfonts}
\usepackage{eucal}
\usepackage{amsmath}
\usepackage{amssymb}
\usepackage{theorem}
\usepackage{mathrsfs}
\usepackage{cases}

\renewcommand{\(}{\left ( }
\renewcommand{\)}{\right ) }
\renewcommand{\H}{{\mathcal H}}
\newcommand{\K}{{\cal K}}

\newcommand{\R}{\mathbb {R}}

\newcommand{\Z}{\mathbb {Z}}

\def\i<#1>{\langle #1 \rangle}
\def\l<#1>{\left\langle #1 \right\rangle}

  \makeatother

  \newtheorem{Theorem}{Theorem}[section]
  
  \newtheorem{Proposition}[Theorem]{Proposition}

\makeatletter

\def\@thesis{}
 \def\id#1{\def\@id{#1}}
 \def\department#1{\def\@department{#1}}

\def\@maketitle{
 \begin{center}
 {\LARGE \@title \par}%
 \vspace{5mm}
 {\Large \@author \par}%
\vspace{5mm}

\end{center}
 \par\vskip 1.5em
 }

\makeatother

\title{{\bf Weak limit theorem for a one-dimensional split-step 
quantum walk}}
  \author{Toru Fuda\footnote{
Department of Mathematics and Science, 
School of Science and Engineering, 
Kokushikan University, 
4-28-1, Setagaya, Setagaya-Ku, Tokyo 154-8515, 
 Japan,
  		\\
E-mail: fudat@kokushikan.ac.jp
},\ \ 
  	Daiju Funakawa\footnote{
Department of Electronics and Information Engineering, Hokkai-Gakuen University, Sapporo 062-8605
 Japan,
  		\\
E-mail: funakawa@hgu.jp
},\ \ 
  	Akito Suzuki\footnote{
Division of Mathematics and Physics, 
Faculty of Engineering, Shinshu University, 
Nagano 380-8553, 
Japan, 
E-mail: akito@shinshu-u.ac.jp}}

\begin{document}
	
\maketitle
	
\begin{abstract}
This paper proves a weak limit theorem 
for a one-dimensional split-step quantum walk
and investigates the limit density function. 
In the density function, 
the difference between two Konno's functions appears. 
\end{abstract}

\section{Introduction}

A large amount of work has been devoted to 
the study of discrete-time quantum walks,
which are viewed as quantum counterparts of
random walks (see \cite{Am03, Ko08, VA15} and references therein). 
One of the most interesting topics in quantum walks 
is a  weak limit theorem,
which was first proved by Konno \cite{Ko02, Ko05}
for a homogeneous quantum walk on $\mathbb{Z}$
and which was extended to more general situations 
by many authors 
\cite{GrJaSc04, IKS05, KoMa10, Li, LiPe09,MiKaKo, SeKo}
(see also \cite{ST12}). 
The weak limit theorem has also been proved 
for quantum walks on the half line \cite{LiPe13},
trees \cite{ChHaKoSe},  joined half lines \cite{ChKoSe},
higher-dimensional lattices \cite{FGMB11, MC, SBJ, WKKK},
crystal lattices \cite{HKSS14}, and several graphs \cite{HS15, MaSe15}.
A random environment case and temporally inhomogeneous case 
were studied in \cite{Ko09, Ma13, MK10}. 
Recently, the weak limit theorem for a nonlinear quantum walk
was established in \cite{MS4a, MS4b}. 

A spatially inhomogeneous discrete-time quantum walk on $\mathbb{Z}$
is described by a unitary evolution operator 
\begin{equation}
\label{0} 
\tilde U = \tilde S C 
\end{equation}
of a product of a shift operator $\tilde S$ and a coin operator $C$
on the Hilbert space 
$\mathcal{H} = \ell^2(\mathbb{Z};\mathbb{C}^2)$.  
Here the shift operator 
$\tilde S$ is defined as
$\tilde S = L \oplus L^*$,
where $\mathcal{H}$ is identified with 
$\ell^2(\mathbb{Z}) \oplus \ell^2(\mathbb{Z})$
and $L$ is the left-shift on $\ell^2(\mathbb{Z})$. 
The coin operator $C$ depends on a position $x \in \mathbb{Z}$
and it is defined as the multiplication operator 
by a family of unitary matrices 
$\{ C(x) \}_{x \in \mathbb{Z}} \subset U(2)$. 
In particular,
in the case where 
$C(x) = C^\prime$ ($x \in \mathbb{Z} \setminus \{0\}$) and
$C(x) = C(0)$ ($x=0$) with $C^\prime, C(0) \in U(2)$, 
the quantum walk is called a one-defect model,
for which the weak limit theorem was proved 
in \cite{Enetal16, EnKo14,KoLuSe13}.  
In \cite{Su15}, 
the weak limit theorem was extended to
the short-range case, where $C_0 := \lim_{|x| \to \infty} C(x)$
and 
\begin{equation}
\label{1}
\|C(x) - C_0\| = O(|x|^{-1-\epsilon}).
\end{equation} 
It is clear that \eqref{1} covers the homogeneous case
with $C(x) \equiv C_0$. 
In the case where 
$C(x) = C_+$ ($x > 0$) and $C(x) = C_-$ ($x < 0$)
with $C_+$, $C_- \in U(2)$,
the quantum walk is called a two-phase model,
for which the weak limit theorem was proved in \cite{EnKoOb15}. 
An anisotropic quantum walk was introduced in \cite{RST1},
where $C(x)$ is assumed to satisfy 
\[ \|C(x) - C_\pm\| = O(|x|^{-1-\epsilon})
	\quad \mbox{for $\pm x > 0$}. \] 
This condition covers the two phase model and 
allows us to prove  
the weak limit theorem \cite{RST2}. 

In this paper, 
we consider 
a split-step quantum walk on $\mathbb{Z}$
introduced in previous papers \cite{FFSmdloc, FFS1dloc},
whose evolution operator is given by $U = SC$. 
The shift operator $S$ of the split-step quantum walk is given by
\begin{equation}
\label{2}
S = \begin{pmatrix} p & q L \\ \bar{q} L^* & -p \end{pmatrix},
\end{equation}
where 
$p \in \mathbb{R}$ and $q \in \mathbb{C}$
satisfy $p^2 + |q|^2 = 1$. 
We suppose that $C(x)$ satisfies \eqref{1},
prove the weak limit theorem, and calculate 
the limit density function.  
The difference between Konno's functions \cite{Ko02} appears 
in the limit density function. 
As put into evidence in \cite{Su15, RST2},
Konno's function always appears in the limit density function
for the one-dimensional quantum walk with the evolution 
$\tilde U = \tilde S C$. 
However, to the authors' best Knowledge,  
this is the first work where
the difference of Konno's functions appears.  

This paper is organized as follows. 
In Section 2, we compare
the split-step quantum walk with the other models. 
In Section 3, we present the weak limit theorem 
for the split-step quantum walk.  
We give the proof in the final section.

\section{Split-step quantum walk and the other walks}
As pointed out in \cite{FFS1dloc}, 
the split-step quantum walk unifies
the spatially inhomogeneous quantum walk 
described by \eqref{0}
and Kitagawa's quantum walk \cite{Ki}. 
Indeed,  if we take $p = 0$, then $S = \tilde S \sigma_1$
and hence $U$ becomes 
the evolution of a spatially inhomogeneous quantum walk
\eqref{0} with a different coin operator. 
If we take $p = \sin (\theta/2)$ and $q = \cos (\theta/2)$, 
then $S = \sigma_1 S_- R(\theta) S_+$,
where $S_+= L^* \oplus 1$, $S_- = 1 \oplus L$,
and $R(\theta)$ is a rotation matrix. 
Taking $C(x) = R(\theta^\prime) \sigma_1$,
we get  
\[ U = \sigma_1 U_{\rm ss}(\theta^\prime, \theta) \sigma_1, \]
where 
$U_{\rm ss}(\theta^\prime, \theta) 
= S_- R(\theta) S_+ R(\theta^\prime)$ 
is the evolution of Kitagawa's split-step quantum walk \cite{Ki}. 
Thus the split-step quantum walk is unitarily equivalent 
to Kitagawa's one. 
For more details, the reader can consult
\cite[Examples 2.1--2.2]{FFS1dloc}. 

The split-step quantum walk 
is viewed as a lazy quantum walk. 
Let $\Psi_0 \in \mathcal{H}$ be 
the initial state and $\Psi_t = U^t \Psi_0$
the state of walker at time $ t = 0, 1, 2, \ldots $.
Then the state evolution is given by 
\begin{equation*}
\Psi_{t+1}(x) = P(x+1)\Psi_t(x+1) + Q(x-1)\Psi_t(x-1) + R(x)\Psi_t(x), 
	\quad x \in \mathbb{Z}
\end{equation*}
with some $2 \times 2$ matrices $P(x)$, $Q(x)$, and $R(x)$. 
The existence of the third term in the left-hand side indicates that
the split-step quantum walk is lazy. 
In \cite{IKS05, LMZZ15},
a lazy quantum walk on $\mathbb{Z}$ is defined 
as three-state quantum walk 
on the Hilbert space $\ell^2(\mathbb{Z};\mathbb{C}^3)$,
whereas the split-step quantum walk is defined as 
a two-state quantum walk on 
$\mathcal{H} = \ell^2(\mathbb{Z};\mathbb{C}^2)$.

\section{Weak limit theorem}
In quantum walks, the position $X_t$ of a walker at time $t$
with an initial state $\Psi_0 \in \H$ ($\|\Psi_0\|=1$) is 
a random variable with the distribution
\[ P(X_t = x) = \|\Psi_t(x)\|^2, \quad x \in \mathbb{Z}, \]
where $\Psi_t := U^t \Psi_0$ is the state of the walker at $t$
with the evolution $U = SC$ and the shift operator $S$ defined in \eqref{2}.  
As shown in \cite{Su15, RST2} for the short-range cases,
$X_t/t$ converges in law to a random variable $V$ as $t \to \infty$. 
This assertion is called the weak limit theorem. 
To show the weak limit theorem,
we suppose the following. 
\begin{flushleft} 
(A.1)\quad There exists a matrix $C_0 \in U(2)$
such that
\[ \|C(x) - C_0 \| \leq \kappa |x|^{-1-\epsilon},
	\quad x \in \mathbb{Z}\setminus \{0\} \]
with some $\kappa, \epsilon > 0$ independent of $x$. 
\end{flushleft} 
For notational simplicity, we assume that 
\begin{flushleft}
(A.2)\quad 
$p, q > 0$, 
\quad  
$\displaystyle C_0 = \begin{pmatrix} a & b \\ b & - a \end{pmatrix},
	\quad a, b > 0$. 
\end{flushleft}
Let $U_0 = SC_0$ be an homogeneous evolution,
whose coin operator is the limit of $C(x)$ at spatial infinity. 
The assumption (A.1) ensures that
the wave operator $W = \mbox{s-}\lim_{t \to \infty} U^{-t} U_0^t$
exists and 
$W^* = \mbox{s-}\lim_{t \to \infty} U_0^{-t} U^t \Pi_{\rm ac}(U)$.  
Here $\Pi_{\rm ac}(U)$ is the projection onto 
the subspace of absolute continuity of $U$.
Moreover, (A.1) allows us to show that 
$U$ has no singular continuous spectrum (see \cite{AsBoJo15, RST1}).
Let $\Pi_{\rm p}(U)$ be the projection 
onto the direct sum of all eigenspaces of $U$
and $f_K(v;r)$ be Konno's function defined as
\[ f_K(v;r) = \frac{\sqrt{1-r^2}}{\pi(1-v^2) \sqrt{r^2 - v^2}}
		 I_{(-r, r)}(v), 
		 \quad v \in \mathbb{R}, \ r > 0, \]
where $I_A$ is the characteristic function of a set $A \subset \R$. 
We use $F$ to denote the Fourier transform form, 
which maps $\mathcal{H}$ to $\K = L^2(\mathbb{T};\mathbb{C}^2;dk/(2\pi))$. 
Let
\begin{equation}
\label{U0} \hat U_0(k) =  
\begin{pmatrix} p & q e^{ik} \\ q e^{-ik} & -p \end{pmatrix} C_0,
	\quad k \in \mathbb{T}  
\end{equation}
and use $u_j(k)$ to denote its normalized 
eigenvector corresponding to eigenvalues 
\[ 
\lambda_j(k) = \exp \left((-1)^{j+1} i \arccos \tau(k) \right),
\] 
where $\tau(k) = pa + qb \cos k$ ($k \in \mathbb{T}$). 
We are now in a position to state our main result. 
\begin{Theorem}\label{thm:2}
Let $X_t$ be as above. Then $X_t/t$ converges in law
to a random $V$, whose distribution is given by
\[  \mu_V(dv)
	= w_0  \delta_0(dv)
		+  w_+(v) f_+(v)dv + w_-(v) f_-(v)dv, \]
where $w_0 = \|\Pi_{\rm p}(U) \Psi_0\|^2$ 
is a nonnegative constant
and nonnegative functions $f_{\pm}(v)$ and $w_\pm(v)$ are given by
    \begin{eqnarray}
f_\pm(v) 
	&=& \frac{\left| f_K(v;q) \mp f_K(v;b) \right|}{2}
		 I_{(-q, q)\cap (-b,b)}(v), \label{f_pm}\\
 w_{\pm}(v) &=&
\begin{cases}
    w_1(2\pi - \arccos g_{\pm}(v))+w_2(\arccos g_{\pm}(v)), & v \geq 0, \\
    w_1(\arccos g_{\pm}(v))+w_2(2\pi - \arccos g_{\pm}(v)), & v < 0 
\end{cases}\label{w_pm}
\end{eqnarray}
with 
\begin{equation}
\label{eqgpm}
g_{\pm}(v)
=\frac{pav^2\pm \sqrt{(q^2-v^2)(b^2-v^2)}}{qb(1-v^2)}
\end{equation} 
and 
\begin{equation}
\label{eqwj}
w_j(k) = \langle u_j(k), (F W^*\Psi_0)(k) \rangle \ (j=1, 2).
\end{equation} 
\end{Theorem}
This theorem says that if $b \not=q$,
we have the difference 
between two Konno's functions 
$f_K(v;q)$ and  $f_K(v;b)$ in the density function. 
To the best of the authors' knowledge, 
this is the first work where such a difference appears. 
 
\section{Proof of the weak limit theorem}
In this section, we prove Theorem \ref{thm:2}. 
The proof proceeds along the same lines as the proof 
of the weak limit theorem in \cite{Su15, RST2}. 
Throughout of this section, we assume (A.1) and (A.2).  

We first remark that 
the Fourier transform $F: \H \to \K$ is unitary and satisfies 
\[ (F\Psi)(k) = \sum_{x \in \Z} e^{-ikx} \Psi(x),
	\quad k \in \mathbb{T}  \]
for all $\Psi \in \mathcal{H}$ with a finite support. 
The Fourier transform $FU_0F^*$ of $U_0$  
is the multiplication operator on $\K$ by 
$\hat U_0(k)$ defined in \eqref{U0}.  
The velocity operator $\hat v_0$
\cite{GrJaSc04, Su15} for the homogeneous evolution $U_0$
is defined so that the Fourier transform $F \hat v_0 F^*$ is given by the multiplication operator on $\K$ by 
the $2\times 2$ hermitian matrix 
\begin{equation}
\label{eqv0} 
(F \hat v_0 F^*)(k)
	= \sum_{j=1,2} v_j(k) |u_j(k)\rangle \langle u_j(k)|,
		\quad k \in \mathbb{T} 
\end{equation} 
with
\begin{equation}
\label{Eq_vj}
v_j(k) = \frac{ (-1)^{j+1}}{ \sqrt{1-\tau^2(k)}} \frac{d \tau}{dk}. 
\end{equation}
The velocity operator $\hat v$ for the inhomogeneous evolution $U$
is defined as $\hat v = W \hat v_0 W^*$. 
We use $E_{\hat v}(\cdot)$ to denote 
the spectral measure of $\hat v$.  
The following proposition can be proved in a way similar to 
\cite[Corollary 2.4]{Su15}. 
\begin{Proposition}\label{wlt}
Suppose (A.1) and (A.2). 
Let $V$ be a random variable with the distribution 
\[ \mu_V(dv) =\|\Pi_{\rm p}(U) \Psi_0\|^2 \delta_0(dv)
	+ \|E_{\hat v}(dv) \Pi_{\rm ac}(U)\Psi_0\|^2. \]
Then, $X_t/t$ converges to $V$ in law as $t \to \infty$. 
\end{Proposition}

It suffices from Proposition \ref{wlt}
to calculate the density function of 
the density function of the continuous part
$\|E_{\hat v}(dv) \Pi_{\rm ac}(U)\Psi_0\|^2$.
To this end, we calculate the Fourier transform of
$\|E_{\hat v}(\cdot) \Pi_{\rm ac}(U)\Psi_0\|^2$. 
Because 
$E_{\hat v}(v)\Pi _{\rm ac}(U) = WE_{\hat v_0}(v)W^*$ and 
$W:\mathcal{H} \to {\rm Ran} \Pi_{\rm ac}(U)$ is unitary,
$\| E_{\hat v}(\cdot)\Pi _{\rm ac}(U)\Psi _0\|^2
=\| E_{\hat v_0}(\cdot)W^*\Psi _0\|^2$.
Combining this with \eqref{eqwj} and \eqref{eqv0}, 
we have
\begin{align}
\notag
 	\int _{-\infty}^\infty e^{i\xi v}d\| E_{\hat{v}}(v)\Pi _{\rm ac}(U)\Psi _0\|^2
 	& =\i< W^*\Psi _0, e^{i\xi \hat v_0}W^*\Psi _0 > \\
    &=\int_{\mathbb{T}}\frac{dk}{2\pi}
    \left(e^{i\xi v_1(k)} w_1(k) + e^{-i\xi v_1(k)}w_2(k)\right), \label{int_dk}
\end{align}
where we have used $v_1(k) = - v_2(k)$ in the last equation.  
In what follows, 
we make the substitutions $v = v_1(k)$ 
in RHS of (\ref{int_dk}). 
To do so, 
we calculate the inverse function of $v = v_1(k)$ and the Jacobian 
$\frac{1}{2\pi}\frac{dk}{dv}$.  
By (\ref{Eq_vj}), we obtain 
\begin{eqnarray*}
\frac{dv}{dk}=\frac{ap( \tau -\frac{a}{p}) ( \tau -\frac{p}{a}) }{\( 1-\tau ^2 \) ^{\frac{3}{2}}}.
\end{eqnarray*}
For the moment, we assume $p>a$. 
Because $\frac{a}{p} < 1 < \frac{p}{a}, \tau < 1$ and $0 < \frac{aq}{bp} < 1$,
the condition $\tau = \frac{a}{p}$ is equivalent to the following condition:
\begin{eqnarray*}
    k = \arccos \frac{aq}{bp}\quad \text{or} \quad
        k = 2\pi - \arccos \frac{aq}{bp}, 
\end{eqnarray*}
where $0 < \arccos \frac{aq}{bp} < \pi$.
By these facts, we have
\begin{eqnarray*}
    \begin{cases}
        \frac{dv}{dk} < 0, & k \in [0, \arccos \frac{aq}{bp})\cup (2\pi - \arccos \frac{aq}{bp}, 2\pi),\\
        \frac{dv}{dk} > 0, & k \in (\arccos \frac{aq}{bp}, 2\pi - \arccos \frac{aq}{bp}).
    \end{cases}
\end{eqnarray*}
Therefore, the function $k \mapsto v$ is injective on each domain of 
$[0, \arccos \frac{aq}{bp}), [\arccos \frac{aq}{bp}, \pi), [\pi, 2\pi - \arccos \frac{aq}{bp})$
and 
$[2\pi - \arccos \frac{aq}{bp}, 2\pi)$. 
Observe that
$v\left([0, \arccos \frac{aq}{bp})\right) = v\left([\arccos \frac{aq}{bp}, \pi)\right) = [-q, 0]$ 
and 
$v\left([\pi, 2\pi - \arccos \frac{aq}{bp})\right) = v\left([2\pi - \arccos \frac{aq}{bp}, 2\pi)\right) =[0,q]$.
Because $v^2=(\tau ')^2\( 1-\tau ^2 \) ^{-1}$,
we know that
$\tau (k)=\frac{pa\pm \sqrt{\( q^2-v^2 \)\( b^2-v^2 \)}}{1-v^2}$.
Since $\tau (k)=pa+qb\cos k$, we obtain
\begin{eqnarray*}
	k&=&\begin{cases}
	\arccos g_+(v),& k\in [0,\ \arccos \frac{aq}{pb}),\\
	\arccos g_-(v),& k\in [\arccos \frac{aq}{pb},\ \pi),\\
	2\pi -\arccos g_-(v), & k\in [\pi,\ 2\pi -\arccos \frac{aq}{pb}),\\
	2\pi -\arccos g_+(v), & k\in [2\pi -\arccos \frac{aq}{pb},\ 2\pi),
\end{cases}
\end{eqnarray*}
where $g_{\pm}(v)$
has been defined in \eqref{eqgpm}. 
By direct calculation, we have
\begin{eqnarray*}
    \frac{d}{dv}\arccos g_{\pm}(v)
= \pm 2\pi \operatorname{sgn}(v)f_{\pm}(v), 
\end{eqnarray*}
where $f_{\pm}(v)$ has been defined in \eqref{f_pm}.
Hence,
\begin{eqnarray}
\frac{1}{2\pi}\frac{dk}{dv}=\begin{cases}
	-f_+(v),&k\in [0,\ \arccos \frac{aq}{pb})\cup [2\pi -\arccos \frac{aq}{pb},\ 2\pi),\\
	f_-(v),& k\in [\arccos \frac{aq}{pb},\ 2\pi -\arccos \frac{aq}{pb}).\label{jacobian}
\end{cases}
\end{eqnarray}
Substituting $v = v_1(k)$ with (\ref{jacobian}), 
we have
\begin{eqnarray*}
    \text{RHS of (\ref{int_dk})} &=& 
        \sum_{\# \in \{+, -\}}\int_{-q}^0dvf_{\#}(v)\left\{e^{i\xi v}w_1(\arccos g_{\#}(v)) +e^{-i\xi v}w_2(\arccos g_{\#}(v))\right\}\\
    &+& \sum_{\# \in \{+, -\}}\int_{0}^qdvf_{\#}(v)\left\{e^{i\xi v}w_1(2\pi - \arccos g_{\#}(v)) +e^{-i\xi v}w_2(2\pi - \arccos g_{\#}(v))\right\}
\\
&=& \int_{-\infty}^{\infty}e^{i\xi v}\left(f_+(v)w_+(v)+f_-(v)w_-(v)\right)dv, 
\end{eqnarray*}
where $w_{\pm}$ has been defined in \eqref{w_pm}.
This completes the proof for the case of $p>a$. 
The same proof works for $p<a$. 
In the case of $p=a$, the proof is immediate, 
because $dv/dk = ap(1-\tau)^{2}/(1+\tau)^{\frac{3}{2}}\geq 0$, i.e., $v = v_1(k)$ is monotonically increasing.  

\medskip


\noindent
{\bf Acknowledgement}
This work was supported by Grant-in-Aid for Young Scientists (B) (No. 26800054).


\end{document}